# 19. COSMIC BACKGROUND RADIATION

## 19.1. Introduction

The observed cosmic microwave background (CMB) radiation provides strong evidence for the hot big bang. The success of primordial nucleosynthesis calculations (see Sec. 16, "Big-bang nucleosynthesis") requires a cosmic background radiation (CBR) characterized by a temperature $kT \sim 1\,\text{MeV}$ at a redshift of $z \simeq 10^9$. In their pioneering work, Gamow, Alpher, and Herman [1] realized this and predicted the existence of a faint residual relic, primordial radiation, with a present temperature of a few degrees. The observed CMB is interpreted as the current manifestation of the hypothesized CBR.

The CMB was serendipitously discovered by Penzias and Wilson [2] in 1965. Its spectrum is well characterized by a $2.73 \pm 0.01\,\text{K}$ black-body (Planckian) spectrum over more than three decades in frequency (see Fig. 19.1). A non-interacting Planckian distribution of temperature $T_i$ at redshift $z_i$ transforms with the universal expansion to another Planckian distribution at redshift $z_r$ with temperature $T_r/(1+z_r) = T_i/(1+z_i)$. Hence thermal equilibrium, once established (e.g. at the nucleosynthesis epoch), is preserved by the expansion, in spite of the fact that photons decoupled from matter at early times. Because there are about $10^9$ photons per nucleon, the transition from the ionized primordial plasma to neutral atoms at $z \sim 1000$ does not significantly alter the CBR spectrum [3].

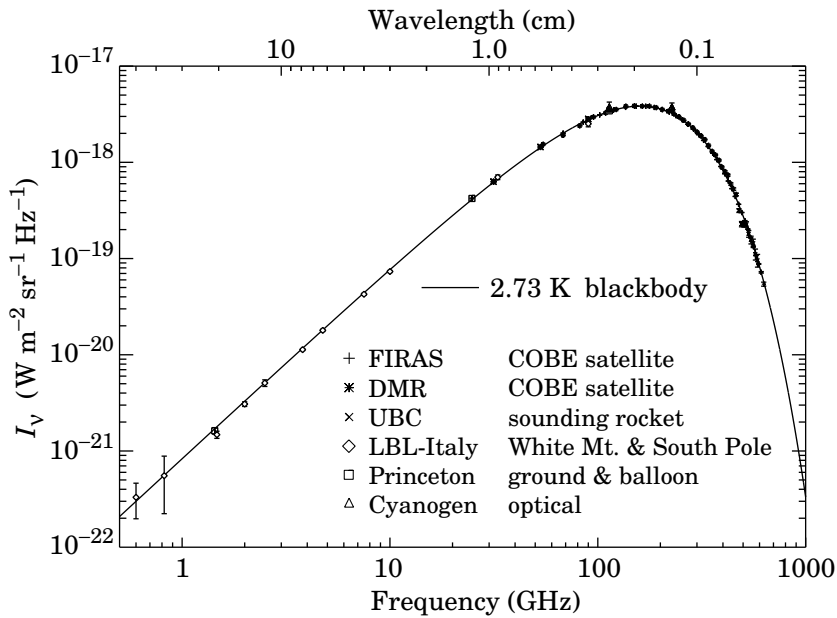

**Figure 19.1:** Precise measurements of the CMB spectrum. The line represents a 2.73 K blackbody, which describes the spectrum very well, especially around the peak of intensity. The spectrum is less well constrained at 10 cm and longer wavelengths. (References for this figure are at the end of this section under "CMB Spectrum References.")

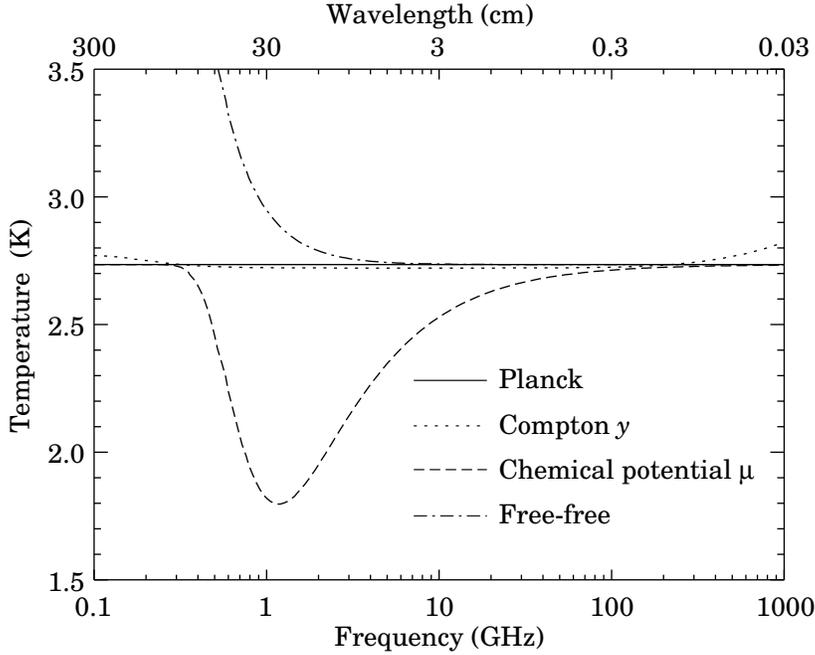

**Figure 19.2:** The shapes of expected, but so far unobserved, CMB distortions, resulting from energy-releasing processes at different epochs.

## 19.2. Theoretical spectral distortions

The remarkable precision with which the CMB spectrum is fitted by a Planckian distribution provides limits on possible energy releases in the early Universe, at roughly the fractional level of $10^{-4}$ of the CBR energy, for redshifts $\lesssim 10^7$ (corresponding to epochs $\gtrsim 1$ year). The following three important classes of spectral distortions (see Fig. 19.2) generally correspond to energy releases at different epochs. The distortion results from the CBR photon interactions with a hot electron gas at temperature $T_e$.

**19.2.1.** *Compton distortion*: Late energy release ($z \lesssim 10^5$). Compton scattering ($\gamma e \to \gamma' e'$) of the CBR photons by a hot electron gas creates spectral distortions by transfering energy from the electrons to the photons. Compton scattering cannot achieve thermal equilibrium for $y < 1$, where

$$y = \int_0^z \frac{kT_e(z') - kT_\gamma(z')}{m_e c^2} \sigma_T\, n_e(z')\, c\, \frac{dt}{dz'}\, dz' \,, \qquad (19.1)$$

is the integral of the number of interactions, $\sigma_T\, n_e(z)\, c\, dt$, times the mean-fractional photon-energy change per collision [4]. For $T_e \gg T_\gamma$ $y$ is also proportional to the integral of the electron pressure $n_e kT_e$ along the line of sight. For standard thermal histories $y < 1$ for epochs later than $z \simeq 10^5$.

The resulting CMB distortion is a temperature decrement

$$\Delta T_{\rm RJ} = -2y\, T_\gamma \qquad (19.2)$$

in the Rayleigh-Jeans ($h\nu/kT \ll 1$) portion of the spectrum, and a rapid rise in temperature in the Wien ($h\nu/kT \gg 1$) region,

*i.e.* photons are shifted from low to high frequencies. The magnitude of the distortion is related to the total energy transfer [4] $\Delta E$ by

$$\Delta E/E_{\mathrm{CBR}} = e^{4y} - 1 \simeq 4y \ . \tag{19.3}$$

A prime candidate for producing a Comptonized spectrum is a hot intergalactic medium. A hot ($T_e > 10^5\,\mathrm{K}$) medium in clusters of galaxies can and does produce a partially Comptonized spectrum as seen through the cluster, known as the Sunyaev-Zel'dovich effect. Based upon X-ray data, the predicted large angular scale total combined effect of the hot intracluster medium should produce $y \lesssim 10^{-6}$ [5].

**19.2.2.** *Bose-Einstein or chemical potential distortion*: Early energy release ($z \sim 10^5$–$10^7$). After many Compton scatterings ($y > 1$), the photons and electrons will reach statistical (not thermodynamic) equilibrium, because Compton scattering conserves photon number. This equilibrium is described by the Bose-Einstein distribution with non-zero chemical potential:

$$n = \frac{1}{e^{x+\mu_0} - 1} \ , \tag{19.4}$$

where $x \equiv h\nu/kT$ and $\mu_0 \simeq 1.4\,\Delta E/E_{\mathrm{CBR}}$, with $\mu_0$ being the dimensionless chemical potential that is required.

The collisions of electrons with nuclei in the plasma produce free-free (thermal bremsstrahlung) radiation: $eZ \to eZ\gamma$. Free-free emission thermalizes the spectrum to the plasma temperature at long wavelengths. Including this effect, the chemical potential becomes frequency-dependent,

$$\mu(x) = \mu_0 e^{-2x_b/x} \ , \tag{19.5}$$

where $x_b$ is the transition frequency at which Compton scattering of photons to higher frequencies is balanced by free-free creation of new photons. The resulting spectrum has a sharp drop in brightness temperature at centimeter wavelengths [6]. The minimum wavelength is determined by $\Omega_B$.

The equilibrium Bose-Einstein distribution results from the oldest non-equilibrium processes ($10^5 < z < 10^7$), such as the decay of relic particles or primordial inhomogeneities. Note that free-free emission (thermal bremsstrahlung) and radiative-Compton scattering effectively erase any distortions [7] to a Planckian spectrum for epochs earlier than $z \sim 10^7$.

**19.2.3.** *Free-free distortion*: Very late energy release ($z \ll 10^3$). Free-free emission can create rather than erase spectral distortion in the late universe, for recent reionization ($z < 10^3$) and from a warm intergalactic medium. The distortion arises because of the lack of Comptonization at recent epochs. The effect on the present-day CMB spectrum is described by

$$\Delta T_{ff} = T_\gamma\, Y_{ff}/x^2, \tag{19.6}$$

where $T_\gamma$ is the undistorted photon temperature, $x$ is the dimensionless frequency, and $Y_{ff}/x^2$ is the optical depth to free-free emission:

$$Y_{ff} = \int_0^z \frac{T_e(z') - T_\gamma(z')}{T_e(z')} \frac{8\pi e^6 h^2 n_e^2\, g}{3 m_e (kT_\gamma)^3\, \sqrt{6\pi\, m_e\, kT_e}} \frac{dt}{dz'} dz' \ . \tag{19.7}$$

Here $h$ is Planck's constant, $n_e$ is the electron density and $g$ is the Gaunt factor [8].

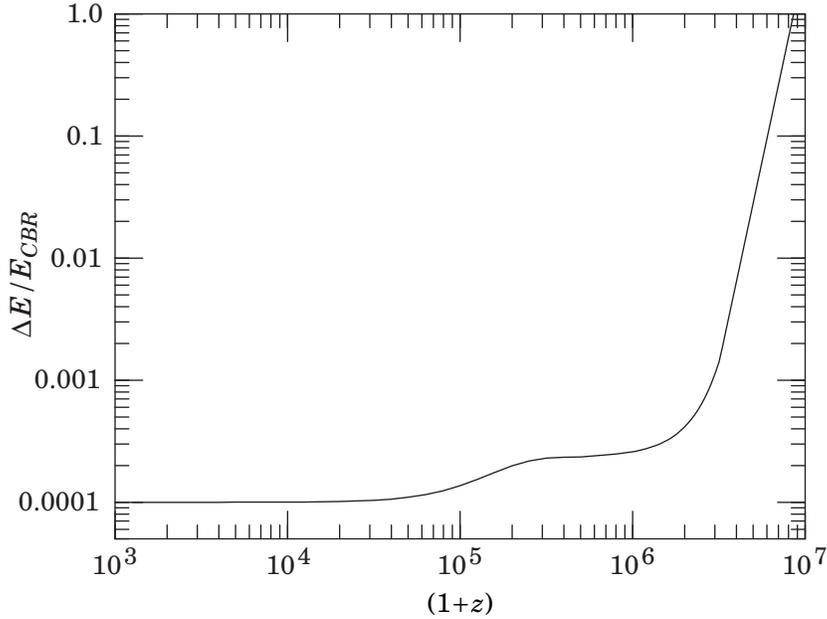

**Figure 19.3:** Upper Limits (95% CL) on fractional energy ($\Delta E/E_{\mathrm{CBR}}$) releases as set by lack of CMB spectral distortions resulting from processes at different epochs. These can be translated into constraints on the mass, lifetime and photon branching ratio of unstable relic particles, with some additional dependence on cosmological parameters such as $\Omega_B$ [9,10].

**19.2.4. *Spectrum summary*:** The CMB spectrum is consistent with a blackbody spectrum over more than three decades of frequency around the peak. A least-squares fit to all CMB measurements yields:

$$T_\gamma = 2.73 \pm 0.01 \text{ K}$$
$$n_\gamma = (2\zeta(3)/\pi^2)T_\gamma^3 \simeq 413 \text{ cm}^{-3}$$
$$\rho_\gamma = (\pi^2/15)T_\gamma^4 \simeq 4.68 \times 10^{-34} \text{ g cm}^{-3} \simeq 0.262 \text{ eV cm}^{-3}$$
$$|y| < 1.5 \times 10^{-5} \quad (95\% \text{ CL})$$
$$|\mu_0| < 9 \times 10^{-5} \quad (95\% \text{ CL})$$
$$|Y_{ff}| < 1.9 \times 10^{-5} \quad (95\% \text{ CL})$$

The limits here [11] correspond to limits [11–13] on energetic processes $\Delta E/E_{\mathrm{CBR}} < 2 \times 10^{-4}$ occurring between redshifts $10^3$ and $5 \times 10^6$ (see Fig. 19.3). The best-fit temperature from the COBE FIRAS experiment is $T_\gamma = 2.728 \pm 0.002$ K [11].

## 19.3. Deviations from isotropy

Penzias and Wilson reported that the CMB was isotropic and unpolarized to the 10% level. Current observations show that the CMB is unpolarized at the $10^{-5}$ level but has a dipole anisotropy at the $10^{-3}$ level, with smaller-scale anisotropies at the $10^{-5}$ level. Standard theories predict anisotropies in linear polarization well below currently achievable levels, but temperature anisotropies of roughly the amplitude now being detected.

It is customary to express the CMB temperature on the sky in a spherical harmonic expansion,

$$T(\theta, \phi) = \sum_{\ell m} a_{\ell m} Y_{\ell m}(\theta, \phi) \,, \tag{19.8}$$

and to discuss the various multipole amplitudes. The power at a given angular scale is roughly $\ell \sum_m |a_{\ell m}|^2 / 4\pi$, with $\ell \sim 1/\theta$.

**19.3.1.** *The dipole*: The largest anisotropy is in the $\ell = 1$ (dipole) first spherical harmonic, with amplitude at the level of $\Delta T/T = 1.23 \times 10^{-3}$. The dipole is interpreted as the result of the Doppler shift caused by the solar system motion relative to the nearly isotropic blackbody field. The motion of the observer (receiver) with velocity $\beta = v/c$ relative to an isotropic Planckian radiation field of temperature $T_0$ produces a Doppler-shifted temperature

$$\begin{aligned} T(\theta) &= T_0 (1 - \beta^2)^{1/2}/(1 - \beta \cos \theta) \\ &= T_0 \left( 1 + \beta \cos \theta + (\beta^2/2) \cos 2\theta + O(\beta^3) \right) \,. \end{aligned} \tag{19.9}$$

The implied velocity [11,14] for the solar-system barycenter is $\beta = 0.001236 \pm 0.000002$ (68% CL) or $v = 371 \pm 0.5 \text{ kms}^{-1}$, assuming a value $T_0 = 2.728 \pm 0.002$ K, towards $(\alpha, \delta) = (11.20^{\text{h}} \pm 0.01^{\text{h}}, -7.0° \pm 0.2°)$, or $(\ell, b) = (264.14° \pm 0.15°, 48.26° \pm 0.15°)$. Such a solar-system velocity implies a velocity for the Galaxy and the Local Group of galaxies relative to the CMB. The derived velocity is $v_{\text{LG}} = 627 \pm 22 \text{ kms}^{-1}$ toward $(\ell, b) = (276° \pm 3°, 30° \pm 3°)$, where most of the error comes from uncertainty in the velocity of the solar system relative to the Local Group.

The Doppler effect of this velocity and of the velocity of the Earth around the Sun, as well as any velocity of the receiver relative to the Earth, is normally removed for the purposes of CMB anisotropy study. The resulting high degree of CMB isotropy is the strongest evidence for the validity of the Robertson-Walker metric.

**19.3.2.** *The quadrupole*: The rms quadrupole anisotropy amplitude is defined through $Q_{\text{rms}}^2/T_\gamma^2 = \sum_m |a_{2m}|^2/4\pi$. The current estimate of its value is $4 \,\mu\text{K} \leq Q_{\text{rms}} \leq 28 \,\mu\text{K}$ for a 95% confidence interval [15]. The uncertainty here includes both statistical errors and systematic errors, which are dominated by the effects of galactic emission modelling. This level of quadrupole anisotropy allows one to set precise limits on anisotropic expansion, shear, and vorticity; all such dimensionless quantities are constrained to be less than about $10^{-5}$.

**19.3.3.** *Smaller angular scales*: The COBE-discovered [16] higher-order ($\ell > 2$) anisotropy is interpreted as being the result of perturbations in the energy density of the early Universe, manifesting themselves at the epoch of the CMB's last scattering. Hence the detection of these anisotropies has provided evidence for the existence of the density perturbations that seeded all the structure we observe today.

In the standard scenario the last scattering takes place at a redshift of approximately 1100, at which epoch the large number of photons was no longer able to keep the hydrogen sufficiently ionized. The optical thickness of the cosmic photosphere is roughly $\Delta z \sim 100$ or about 5 arcminutes, so that features smaller than this size are damped.

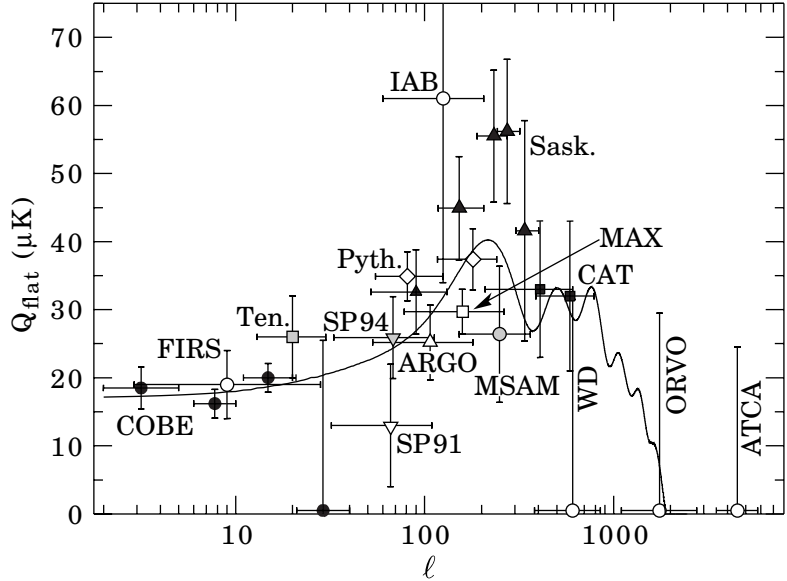

**Figure 19.4:** Current status of CMB anisotropy observations, adapted from Scott, Silk, & White (1995) [17]. This is a representation of the results from COBE, together with a wide range of ground- and balloon-based experiments which have operated in the last few years. Plotted are the quadrupole amplitudes for a flat (unprocessed scale-invariant spectrum of primordial perturbations, *i.e.*, a horizontal line) anisotropy spectrum that would give the observed results for each experiment. In other words each point is the normalization of a flat spectrum derived from the individual experiments. The vertical error bars represent estimates of 68% CL, while the upper limits are at 95% CL. Horizontal bars indicate the range of $\ell$ values sampled. The curve indicates the expected spectrum for a standard CDM model ($\Omega_0 = 1, \Omega_B = 0.05, h = 0.5$), although true comparison with models should involve convolution of this curve with each experimental filter function. (References for this figure are at the end of this section under "CMB Anisotropy References.")

Anisotropies are observed on angular scales larger than this damping scale (see Fig. 19.4), and are consistent with those expected from an initially scale-invariant power spectrum (flat = independent of scale) of potential and thus metric fluctuations. It is believed that the large scale structure in the Universe developed through the process of gravitational instability, where small primordial perturbations in energy density were amplified by gravity over the course of time. The initial spectrum of density perturbations can evolve significantly in the epoch $z > 1100$ for causally connected regions (angles $\lesssim 1° \, \Omega_{tot}^{1/2}$). The primary mode of evolution is through adiabatic (acoustic) oscillations, leading to a series of peaks that encode information about the perturbations and geometry of the universe, as well as information on $\Omega_0$, $\Omega_B$, $\Omega_\Lambda$ (cosmological constant), and $H_0$ [17]. The location of the first acoustic peak is predicted to be at $\ell \sim 220 \, \Omega_{tot}^{-1/2}$ or $\theta \sim 0.3° \, \Omega_{tot}^{1/2}$ and its amplitude increases with increasing $\Omega_B$.

Theoretical models often predict a power spectrum in spherical harmonic amplitudes, since the models lead to primordial fluctuations and thus $a_{\ell m}$ that are Gaussian random fields, and hence the power spectrum in $\ell$ is sufficient to characterize the results. The power at each $\ell$ is $(2\ell + 1)C_\ell/(4\pi)$, where $C_\ell \equiv \langle |a_{\ell m}|^2 \rangle$. For an idealized full-sky observation, the variance of each measured $C_\ell$ is $[2/(2\ell + 1)]C_\ell^2$. This sampling variance (known as cosmic variance) comes about because each $C_\ell$ is chi-squared distributed with $(2\ell + 1)$ degrees of freedom for our observable volume of the Universe [18].

Figure 19.5 shows the theoretically predicted anisotropy power spectrum for a sample of models, plotted as $\ell(\ell + 1)C_\ell$ versus $\ell$ which is the power per logarithmic interval in $\ell$ or, equivalently, the two-dimensional power spectrum. If the initial power spectrum of perturbations is the result of quantum mechanical fluctuations produced and amplified during inflation, then the shape of the anisotropy spectrum is coupled to the ratio of contributions from density (scalar) and gravity wave (tensor) perturbations. If the energy scale of inflation at the appropriate epoch is at the level of $\simeq 10^{16}$GeV, then detection of the effect of gravitons is possible, as well as partial reconstruction of the inflaton potential. If the energy scale is $\lesssim 10^{14}$GeV, then density fluctuations dominate and less constraint is possible.

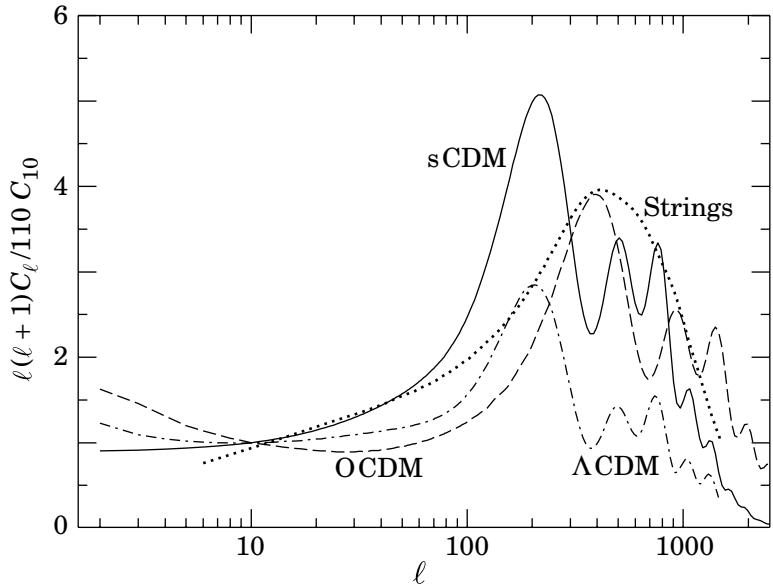

**Figure 19.5:** Examples of theoretically predicted $\ell(\ell + 1)C_\ell$ or CMB anisotropy power spectra. **sCDM** is the standard cold dark matter model with $h = 0.5$ and $\Omega_B = 0.05$. **$\Lambda$CDM** is a model with $\Omega_{\rm tot} = \Omega_\Lambda + \Omega_0 = 1$, with $\Omega_\Lambda = 0.3$ and $h = 0.8$. **OCDM** is an open model with $\Omega_0 = 0.3$ and $h = 0.75$ (see [19] for models). **Strings** is a model where cosmic strings are the primary source of large scale structure [20]. The plot indicates that precise measurements of the CMB anisotropy power spectrum could distinguish between current models.

Fits to data over smaller angular scales are often quoted as the expected value of the quadrupole $\langle Q \rangle$ for some specific theory, *e.g.* a model with power-law initial conditions (primordial density perturbation power spectrum $P(k) \propto k^n$). The full 4-year COBE DMR data give $\langle Q \rangle = 15.3^{+3.7}_{-2.8}$ $\mu$K, after projecting out the slope dependence, while the best-fit slope is $n = 1.2 \pm 0.3$, and for a pure $n = 1$ (scale-invariant potential perturbation) spectrum $\langle Q \rangle (n = 1) = 18 \pm 1.6$ $\mu$K [15,21]. The conventional notation is such that $\langle Q \rangle^2 / T_\gamma^2 = 5C_2/4\pi$. The fluctuations measured by other experiments can also be quoted in terms of $Q_{\text{flat}}$, the equivalent value of the quadrupole for a flat ($n = 1$) spectrum, as presented in Fig. 19.4.

It now seems clear that there is more power at sub-degree scales than at COBE scales, which provides some model-dependent information on cosmological parameters [17,22], for example $\Omega_B$. In terms of such parameters, fits to the COBE data alone yield $\Omega_0 > 0.34$ at 95% CL [23] and $\Omega_{\text{tot}} < 1.5$ also at 95% CL [24], for inflationary models. Only somewhat weak conclusions can be drawn based on the current smaller angular scale data (see Fig. 19.4). A sample preliminary fit [25] finds $\Omega_{\text{tot}} = 0.7^{+1.0}_{-0.4}$ and $30 < H_0 < 70$ km s$^{-1}$ Mpc$^{-1}$ for a limited range of cosmological models.

However, new data are being acquired at an increasing rate, with a large number of improved ground- and balloon-based experiments being developed. It appears that we are not far from being able to distinguish crudely between currently favored models, and to begin a more precise determination of cosmological parameters. A vigorous suborbital and interferometric program could map out the CMB anisotropy power spectrum to about 10% accuracy and determine several parameters at the 10 to 20% level in the next few years. Ultimately, on the scale of a perhaps 5–10 years, there is the prospect of another satellite mission which could provide a precise measurement of the power spectrum down to scales of 10 arcminutes, allowing us to decode essentially all of the information that it contains [26].